\documentclass[12pt]{article}
\usepackage{amsmath}
\usepackage{amssymb}
\usepackage{amsfonts}
\usepackage{graphics}

\numberwithin{equation}{section}

\title{Rational extensions of an oscillator-shaped quantum well potential in a position-dependent mass background}
\author{C Quesne\\ 
{\small Physique Nucl\'eaireTh\'eorique et Physique Math\'ematique,  Universit\'e Libre de Bruxelles,} \\ 
{\small Campus de la Plaine CP229, Boulevard~du Triomphe, B-1050 Brussels, Belgium}\\
{\small E-mail: Christiane.Quesne@ulb.be}}
\date{ }
\begin{document}
\baselineskip=22pt plus 1pt minus 1pt
\maketitle

\begin{abstract} 
We show that a recently proposed oscillator-shaped quantum well model associated with a position-dependent mass can be solved by applying a point canonical transformation to the constant-mass Schr\"odinger equation for the Scarf I potential. On using the known rational extension of the latter connected with $X_1$-Jacobi exceptional orthogonal polynomials, we build a rationally-extended position-dependent mass model with the same spectrum as the starting one. Some more involved position-dependent mass models associated with $X_2$-Jacobi exceptional orthogonal polynomials are also considered.
\end{abstract}

\newpage

\section{Introduction}

Since many years, the Schr\"odinger equation in a position-dependent mass (PDM) background has arisen much interest due to the utmost relevance of the PDM concept in a wide variety of physical situations, such as in electronic properties of semi-conductors and quantum dots, in quantum liquids, $^3$He clusters, and metal clusters, as well as in energy density many-body problems \cite{bastard, weisbuch, serra, harrison, barranco, geller, arias, puente, ring, bonatsos, willatzen, chamel}. Finding exact solutions of such a Schr\"odinger equation is therefore very useful for understanding some physical phenomena and for testing some approximation methods. The interest in such a study is also reinforced by the fact that the PDM Schr\"odinger equation is equivalent \cite{cq04} to the Schr\"odinger equation in curved space \cite{schro, kalnins96, kalnins97} and to that resulting from the use of deformed commutation relations \cite{kempf, hinrichsen, witten}.\par
%
%
Several techniques are available for generating PDM and potential pairs leading to exact solutions for the Schr\"odinger equation (see, e.g., \cite{cq04, bagchi05, cq06}). Among them, one of the most powerful is the point canonical transformation (PCT) applied to an exactly-solvable constant-mass Schr\"odinger equation \cite{bagchi04, cq09a}. Recently, such an approach has proved its efficiency again \cite{cq21, cq22, cq23}.\par
%
%
During recent years, for constant-mass Schr\"odinger equations, there has been much interest in constructing new exactly-solvable rational extensions of well-known quantum wells after the introduction of exceptional orthogonal polynomials (EOPs) \cite{gomez09}. The latter form orthogonal and complete polynomial sets although they admit some gaps in the sequence of their degrees in contrast with classical orthogonal polynomials (COPs). EOPs were indeed shown to be related to the Darboux transformation in the context of shape invariant potentials in supersymmetric quantum mechanics \cite{cq08, cq09b}. Infinite families of shape invariant potentials were then constructed in relation to $X_m$ EOPs \cite{odake09}, as well as generalizations thereof (see, e.g., \cite{gomez12, odake11}).\par
%
%
{}For PDM Schr\"odinger equations, there have been less studies aiming at building rationally-extended potentials related to EOPs. It is however obvious that if, by using a PCT, a PDM Schr\"odinger equation can be derived from a conventional one, whose rational extensions are well known, then the same PCT applied to such extensions may provide some rational extensions of the starting PDM Schr\"odinger equation. Such a procedure was already applied with success to some problems in curved space \cite{cq16}.\par
%
%
The purpose of the present paper is to present an example of application of this method in the PDM context. The starting PDM Schr\"odinger equation will be a model corresponding to an oscillator-shaped quantum well potential, whose eigenvalues and eigenfunctions were recently obtained by Jafarov and Nagiyev by directly solving the equation \cite{jafarov}.\par
%
%
This paper is organized as follows. In section~2, the model of Ref.~\cite{jafarov} is reviewed and shown to be derivable by applying the PCT technique to the constant-mass Scarf I potential. In section~3, rational extensions of the latter related to $X_1$- and $X_2$-Jacobi EOPs are then used to build some rational extensions of the PDM model. Finally, section~4 contains the conclusion.\par
%
%
\section{Oscillator-shaped quantum well potential model and its derivation by the PCT technique}

In \cite{jafarov}, Jafarov and Nagiyev considered the Schr\"odinger equation
\begin{equation}
  \left(-\frac{d}{dx} \frac{1}{M(x)} \frac{d}{dx} + V_{\rm eff}(x)\right) \psi(x) = E \psi(x),  \label{eq:PDM-eqn}
\end{equation}
where $M(x)$ and $V_{\rm eff}(x)$ are defined by\footnote{Note that we have adopted here units wherein $\hbar= 2m_0 = 1$ in the original paper.}
\begin{align}
  & M(x) = \frac{ab}{(x-a)(b-x)}, \qquad V_{\rm eff}(x) = \frac{1}{4} M(x) \omega^2 x^2 = \frac{ab\omega^2 
     x^2}{4(x-a)(b-x)}, \nonumber \\
  &\quad 0 < a < x < b.  \label{eq:mass}
\end{align}
Such a potential is an oscillator-shaped quantum well confined in a cavity between two infinite walls located at $x=a$ and $x=b$.\par
%
%
It is worth observing that, as shown in (\ref{eq:PDM-eqn}), the BenDaniel-Duke form \cite{bendaniel} was adopted in \cite{jafarov} for the kinetic energy operator. This is only a special case of the von Roos general two-parameter form of the latter \cite{vonroos}. Other orderings of the mass and the differential operator, such as the Zhu-Kroemer \cite{zhu} or the Mustafa-Mazharimousavi \cite{mustafa07, mustafa19} ones, might have been chosen, but, as shown elsewhere \cite{cq22}, in general they do not change the results much.\par
%
%
By directly solving equation (\ref{eq:PDM-eqn}), Jafarov and Nagiyev found that the spectrum of the model is given by
\begin{equation}
  E_n = \frac{b+a}{b-a} \omega \left(n+\frac{1}{2}\right) + \frac{1}{ab} n(n+1) + \omega^2 \frac{a^2b^2}
  {(b-a)^2}, \qquad n=0, 1, 2, \ldots,  \label{eq:E}
\end{equation}
with corresponding wavefunctions
\begin{equation}
  \psi_n(x) = N_n (x-a)^{\frac{\omega}{2}\frac{a^2b}{b-a}} (b-x)^{\frac{\omega}{2}\frac{ab^2}{b-a}}
  P_n^{\left(\omega \frac{ab^2}{b-a}, \omega\frac{a^2b}{b-a}\right)}\left(\frac{2x-a-b}{b-a}\right),
  \label{eq:psi}
\end{equation}
\begin{eqnarray}
  N_n & = &\left\{\left(\omega ab\frac{a+b}{b-a}+2n+1\right) n! \Gamma\left(\omega ab\frac{a+b}{b-a}
      +n+1\right)\right\}^{1/2} \nonumber \\
  & & {}\times \left\{(b-a)^{\omega ab\frac{a+b}{b-a}+1} \Gamma\left(\omega\frac{ab^2}{b-a}+n+1\right)
      \Gamma\left(\omega\frac{a^2b}{b-a}+n+1\right)\right\}^{-1/2},  \label{eq:N}
\end{eqnarray}
expressed in terms of Jacobi polynomials $P_n^{(\alpha,\beta)}(z)$ and vanishing at $x=a$ and $x=b$.\footnote{In \cite{jafarov}, there is an additional (optional) phase factor $(-1)^n$.}\par
%
%
These results may be alternatively derived by applying a PCT to the constant-mass Schr\"odinger equation for the Scarf I potential \cite{cq08, cq09b}
\begin{equation}
  \left(-\frac{d^2}{du^2} + U(u)\right) \phi_n(u) = \epsilon_n \phi_n(u),  \label{eq:SE}
\end{equation}
where
\begin{equation}
  U(u) = (A^2 + B^2 - A) \sec^2 u - B(2A-1) \tan u \sec u, \quad -\frac{\pi}{2} < u < \frac{\pi}{2}, \quad
  0 < B < A-1,  \label{eq:U}
\end{equation}
\begin{equation}
  \epsilon_n = (A+n)^2, \qquad n=0, 1, 2, \ldots,  \label{eq:epsilon}
\end{equation}
and 
\begin{align}
  \phi_n(u) &= {\cal N}_n (1-\sin u)^{(A-B)/2} (1+\sin u)^{(A+B)/2} P_n^{\left(A-B-\frac{1}{2}, A+B-\frac{1}{2}
       \right)}(\sin u),  \label{eq:phi} \\
  {\cal N}_n &= \left(\frac{(2A+2n) n!\, \Gamma(2A+n)}{2^{2A}\Gamma\left(A-B+n+\frac{1}{2}\right)
       \Gamma\left(A+B+n+\frac{1}{2}\right)}\right)^{1/2}.  \label{eq:N-bis} 
\end{align}
\par
%
%
A PCT transforming a constant-mass equation such as (\ref{eq:SE}) into a PDM equation of type (\ref{eq:PDM-eqn}) \cite{bagchi04, cq09a} consists in making a change of variable
\begin{equation}
  u(x) = \bar{a} v(x) + \bar{b}, \qquad v(x) = \int^x \sqrt{M(x')} dx',  \label{eq:u-v}
\end{equation}
and a change of function
\begin{equation}
  \phi(u(x)) \propto [M(x)]^{-1/4} \psi(x).  \label{eq:phi-psi}
\end{equation}
Here, $\bar{a}$ and $\bar{b}$ are assumed to be two real parameters. The potential $V_{\rm eff}(x)$, defined on a possibly different interval, and the bound-state energies $E_n$ of the PDM Schr\"odinger equation are given in terms of the potential $U(u)$ and the bound-state energies $\epsilon_n$ of the constant-mass one, by
\begin{equation}
  V_{\rm eff}(x) = \bar{a}^2 U(u(x)) + \frac{M^{\prime\prime}}{4M^2} - \frac{7M^{\prime2}}{16M^3} + \bar{c},
  \label{eq:V-U}
\end{equation}
and
\begin{equation}
  E_n = \bar{a}^2 \epsilon_n + \bar{c},  \label{eq:E-epsilon}
\end{equation}
where a prime denotes derivative with respect to $x$ and $\bar{c}$ is some additional real constant. From (\ref{eq:phi-psi}), the corresponding bound-state wavefunctions are given by
\begin{equation}
  \psi_n(x) = \lambda [M(x)]^{1/4} \phi_n(u(x)) \label{eq:psi-phi}
\end{equation}
provided they are normalizable on the defining interval of $x$. Here $\lambda$ is some constant that may arise from the change of normalization when going from $u$ to $x$.\par
%
%
{}For the mass chosen in (\ref{eq:mass}), one finds that the mass-dependent term in (\ref{eq:V-U}) is given by
\begin{equation}
  \frac{M^{\prime\prime}}{4M^2} - \frac{7M^{\prime2}}{16M^3} = \frac{1}{4ab} \left(1+\frac{(b-a)^2}
  {4(x-a)(b-x)}\right)
\end{equation}
and the change of variable (\ref{eq:u-v}) gives
\begin{equation}
  u(x) = \bar{a} \sqrt{ab} \arcsin \frac{2x-a-b}{b-a} + \bar{b}.
\end{equation}
On assuming
\begin{equation}
  \bar{a} = -\frac{1}{\sqrt{ab}}, \qquad \bar{b} = 0,  \label{eq:a-bar}
\end{equation}
one gets
\begin{equation}
  u(x) =- \arcsin\frac{2x-a-b}{b-a} \qquad \text{or} \qquad \sin u = - \frac{2x-a-b}{b-a}.
\end{equation}
\par
%
%
{}From the latter and (\ref{eq:phi}), one finds that equation (\ref{eq:psi-phi}) for the wavefunctions amounts to (\ref{eq:psi}), provided one assumes
\begin{equation}
  A = \frac{1}{2}\left(\omega\frac{ab}{b-a}(a+b) +1\right), \qquad B = \frac{1}{2}\omega ab,  \label{eq:A-B}
\end{equation}
which leads to the condition $2\omega a^2 b>b-a$, and the normalization factors in (\ref{eq:N}) and (\ref{eq:N-bis}) are related by
\begin{equation}
  N_n = \lambda {\cal N}_n (ab)^{1/4} \left(\frac{2}{b-a}\right)^{\frac{1}{2}\left(\omega \frac{ab}
  {b-a}(a+b)+1\right)}.  \label{eq:N-N-bis}
\end{equation}
The corresponding eigenvalues (\ref{eq:E-epsilon}) reduce to (\ref{eq:E}) provided one chooses
\begin{equation}
  \bar{c} = - \frac{\omega^2}{4} ab - \frac{1}{4ab}.  \label{eq:c-bar}
\end{equation}
Finally, one may check that for the choice of parameters made in (\ref{eq:a-bar}), (\ref{eq:A-B}), and (\ref{eq:c-bar}), potential (\ref{eq:V-U}) is indeed given by (\ref{eq:mass}), as it should be.\par
%
%
The only point that remains to be checked is that the normalization factor $N_n$, as given by (\ref{eq:N-N-bis}), reduces to (\ref{eq:N}). This is straightforward because the factor $\lambda$ coming from the change of normalization from $-\pi/2 < u < \pi/2$ to $a <x <b$ is easily shown to be given by $\lambda = (ab)^{-1/4}$.\par
%
%
\section{Rational extensions of the PDM model}

\setcounter{equation}{0}

The simplest rational extension of the Scarf I potential can be expressed as \cite{cq09b}
\begin{equation}
  U_{\rm ext}(u) = U(u) + U_{\rm rat}(u),  \label{eq:Scarf-ext}
\end{equation}
where $U(u)$ is given in (\ref{eq:U}) and 
\begin{equation}
  U_{\rm rat}(u) = \frac{2(2A-1)}{2A-1-2B\sin u} - \frac{2[(2A-1)^2-4B^2]}{(2A-1-2B\sin u)^2}.
\end{equation}
It has the same spectrum (\ref{eq:epsilon}) as $U(u)$ and the corresponding wavefunctions are given by
\begin{equation}
  \phi_n(u) = {\cal N}_n \frac{(1-\sin u)^{\frac{1}{2}(A-B)}(1+\sin u)^{\frac{1}{2}(A+B)}}{2A-1-2B\sin u}
  \hat{P}_{n+1}^{\left(A-B-\frac{1}{2}, A+B-\frac{1}{2}\right)}(\sin u),
\end{equation}
with
\begin{align}
  {\cal N}_n &= \frac{B}{2^{A-2}} [(2A+2n)n!\, \Gamma(2A+n)]^{1/2} \left[\left(A-B+n+\frac{1}{2}\right)
       \left(A+B+n+\frac{1}{2}\right)\right]^{-1/2} \nonumber \\
  &\quad \times \left[\Gamma\left(A-B+n-\frac{1}{2}
      \right)\Gamma\left(A+B+n-\frac{1}{2}\right)\right]^{-1/2}.  \label{eq:cal-N}
\end{align}
Here $\hat{P}_{n+1}^{(\alpha,\beta)}(z)$ denotes an $(n+1)$th-degree $X_1$-Jacobi EOP, as defined in Ref.~\cite{gomez09}. For $n=0$, 1, 2, \ldots, such polynomials are known to form an orthogonal and complete set with respect to the positive-definite measure $(1-x)^{\alpha} (1+x)^{\beta} \left(x - \frac{\beta+\alpha}{\beta-\alpha}\right)^{-2} dx$. It is worth noting here that some other notations for them are found in the literature, for instance $\hat{P}^{(\alpha,\beta)}_{1,n+1}(z) = (\alpha+n)(\beta-\alpha)(\alpha+n+1)^{-1} \hat{P}^{(\alpha,\beta)}_{n+1}(z)$ in Refs.~\cite{gomez13, liaw, bonneux}.\par
%
%
On replacing $U(u)$ by $U_{\rm ext}(u)$ in (\ref{eq:V-U}) and keeping the same values for the parameters as in section~2, we obtain that $V_{\rm eff}(x)$ is replaced by
\begin{equation}
  V_{\rm eff, ext}(x) = V_{\rm eff}(x) + V_{\rm eff, rat}(x),  \label{eq:V-ext}
\end{equation}
with $V_{\rm eff}(x)$ given in (\ref{eq:mass}), while
\begin{equation}
  V_{\rm eff, rat}(x) = \frac{1}{abx^2}[(a+b)x - 2ab].
\end{equation}
The resulting potential (\ref{eq:V-ext}) is still an oscillator-shaped quantum well confined between two infinite walls at $x=a$ and $x=b$. Its minimum, however, which was located at $x_{\rm min} = 2ab/(a+b)$, is slightly displaced to the left because $V_{\rm eff, rat}(x)$ is negative or positive according to whether $x<x_{\rm min}$ or $x>x_{\rm min}$. As an example, we show in figure~1 the potential $V_{\rm eff, ext}(x)$ corresponding to $\omega=a=1$, $b=3$, and given by
\begin{equation}
  V_{\rm eff, ext}(x) = \frac{3x^2}{4(x-1)(3-x)} + \frac{4x-6}{3x^2}.  \label{eq:pot-ext}
\end{equation}
\par
%
%
The spectrum of $V_{\rm eff, ext}(x)$ remains the same as given by (\ref{eq:E}), but the wavefunctions become
\begin{eqnarray}
  \psi_n(x) & = & N_n \frac{1}{x} (x-a)^{\frac{\omega}{2}\frac{a^2b}{b-a}} (b-x)^{\frac{\omega}{2}
      \frac{ab^2}{b-a}} \hat{P}_{n+1}^{\left(\omega \frac{ab^2}{b-a},\omega \frac{a^2b}{b-a}\right)}
      \left(\frac{2x-a-b}{b-a}\right), \nonumber \\
   & & n=0, 1, 2, \ldots,
\end{eqnarray}
where
\begin{equation}
  N_n = \frac{\lambda}{\omega} (ab)^{-3/4} \left(\frac{2}{b-a}\right)^{\frac{1}{2}\left(\omega ab\frac{a+b}{b-a}
  -1\right)} {\cal N}_n
\end{equation}
with ${\cal N}_n$ expressed in (\ref{eq:cal-N}) and $A$, $B$ given in (\ref{eq:A-B}). With $\lambda$ given by $\lambda = (ab)^{-1/4}$ again, $N_n$ can be written as
\begin{align}
  N_n &= \left\{\left(\omega ab\frac{a+b}{b-a}+2n+1\right) n!\, \Gamma\left(\omega ab\frac{a+b}{b-a}+n+1
      \right)\right\}^{1/2} \nonumber \\
  &\quad \times \left\{(b-a)^{\omega ab\frac{a+b}{b-a}-1} \left(\omega \frac{ab^2}{b-a}+n+1\right)
      \left(\omega \frac{a^2b}{b-a}+n+1\right)\right\}^{-1/2} \nonumber \\
  &\quad \times \left\{\Gamma\left(\omega\frac{ab^2}{b-a}+n\right) \Gamma\left(\omega\frac{a^2b}{b-a}
      +n\right)\right\}^{-1/2}.
\end{align}
As examples, the first three eigenfunctions $\psi_0(x)$, $\psi_1(x)$, and $\psi_2(x)$ of potential (\ref{eq:pot-ext}), corresponding to $E_0 = \frac{13}{4}$, $E_1 = \frac{71}{12}$, and $E_2 = \frac{37}{4}$, respectively, are displayed in figure~2.\par 
%
%
\begin{figure}
\begin{center}
\includegraphics{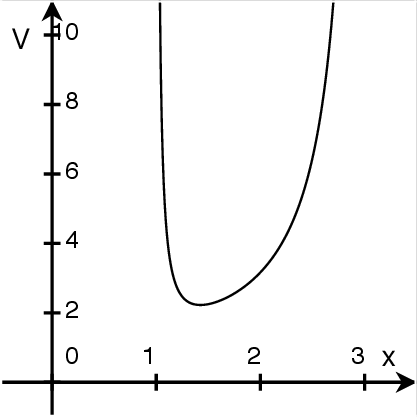}
\caption{Plot of potential $V_{\rm eff,ext}(x)$, defined in (\ref{eq:pot-ext}), in terms of $x$.}
\end{center}
\end{figure}
\par
%
%
\begin{figure}
\begin{center}
\includegraphics{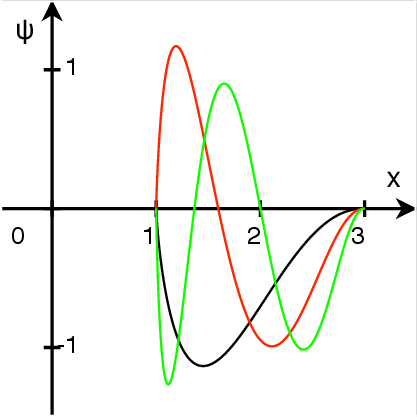}
\caption{Plot of wavefunctions $\psi_0(x)$ (black line), $\psi_1(x)$ (red line), and $\psi_2(x)$ (green line) for potential (\ref{eq:pot-ext}) in terms of $x$.}
\end{center}
\end{figure}
\par
%
%
Let us now sketch more involved extensions of $V_{\rm eff}(x)$, obtained from the rational extensions of Scarf I potential related to $X_2$-Jacobi EOPs \cite{cq09b}. The latter belong to three different types, where in all cases $U_{\rm ext}(u)$ can be written as shown in (\ref{eq:Scarf-ext}) with
\begin{equation}
  U_{\rm rat}(u) = \frac{N_1(u)}{D(u)} + \frac{N_2(u)}{[D(u)]^2}.
\end{equation}
Here
\begin{equation}
\begin{split}
  N_1(u) &= -4\left[(2A-1)(2B-1)(2B-2)\sin u + 2(2A-1)^2 - (2B-2)^2 (2B+1)\right], \\
  N_2(u) &= -8(2B-2)(2A-2B+1)(2A+2B-3) \\
  &\quad \times \left[2(2A-1)(2B-1)\sin u - (2A-1)^2 - 2B(2B-2)\right], \\
  D(u) &= (2B-1) \left[(2B-2)\sin u - (2A-1)\right]^2 - (2A-2B+1)(2A+2B-3),  
\end{split}
\end{equation}
for type I (with $1<B<A-1$),
\begin{equation}
\begin{split}
  N_1(u) &= -4\left[(2A-1)(2B+1)(2B+2)\sin u - 2(2A-1)^2 - (2B+2)^2 (2B-1)\right], \\
  N_2(u) &= 8(2B+2) (2A-2B-3)(2A+2B+1) \\
  &\quad \times\left[2(2A-1)(2B+1)\sin u  - (2A-1)^2 - 2B(2B+2)\right], \\
  D(u) &= (2B+1)\left[(2B+2) \sin u - (2A-1)\right]^2 + (2A-2B-3)(2A+2B+1),
\end{split}
\end{equation}
for type II (with $0<B<A-\frac{3}{2}$), and
\begin{equation}
\begin{split}
  N_1(u) &= -8\left[B(2A-2)(2A-3) \sin u - A(2A-3)^2 + 4B^2\right], \\
  N_2(u) &= 8(2A-3)(2A-2B-3)(2A+2B-3) \\
  &\quad \times\left[4B(2A-2)\sin u - 4B^2 - (2A-1)(2A-3)\right], \\
  D(u) &= (2A-2) \left[(2A-3)\sin u - 2B\right]^2 + (2A-2B-3)(2A+2B-3),
\end{split}
\end{equation}
for type III (with $0<B<A-\frac{3}{2}$).\par
%
%
By proceeding as above, we obtain that in (\ref{eq:V-ext}), $V_{\rm eff,rat}(x)$ is given by
\begin{equation}
  V_{\rm eff,rat}(x) = \frac{{\cal N}_1(x)}{{\cal D}(x)} + \frac{{\cal N}_2(x)}{[{\cal D}(x)]^2},
\end{equation}
where
\begin{equation}
\begin{split}
  {\cal N}_1(x) &= \frac{8}{ab(b-a)^2}\bigl\{\omega ab(\omega ab-1)(\omega ab-2)[(a+b)x-2ab] - \omega^2
       a^2 b^2(a+b)^2 \\
  &\quad - (\omega ab-2)(b-a)^2\bigr\}, \\
  {\cal N}_2(x) &= \frac{128\omega}{(b-a)^4} (\omega ab-2) \big[\omega^2 a^3 b^3 + (\omega ab-1)
      (b-a)^2\big] \\
  &\quad \times \big[(a+b)(\omega ab-1)x - ab (\omega ab-2)\big], \\
  {\cal D}(x) &= \frac{4}{(b-a)^2} \bigl\{(\omega ab-1) [(\omega ab-2)x + 2a][(\omega ab-2)x + 2b]
      - \omega^2 a^3 b^3\bigr\}, 
\end{split}
\end{equation}
for type I (with $\omega ab> \max\left(2, \frac{b-a}{2a}\right)$),
\begin{equation}
\begin{split}
  {\cal N}_1(x) &= \frac{8}{ab(b-a)^2}\bigl\{\omega ab (\omega ab+1)(\omega ab+2)[(a+b)x-2ab]
      + \omega^2 a^2 b^2(a+b)^2 \\
  &\quad - (b-a)^2(\omega ab+2)\bigr\}, \\
  {\cal N}_2(x) &= \frac{128\omega}{(b-a)^4}(\omega ab+2)\big[\omega^2a^3b^3 - (\omega ab+1)(b-a)^2
      \big] \\
  &\quad \times \big[-(a+b)(\omega ab+1)x + ab(\omega ab+2)\big], \\
  {\cal D}(x) &= \frac{4}{(b-a)^2}\bigl\{(\omega ab+1)\big[(\omega ab+2)x-2a\big]\big[(\omega ab+2)x-2b
      \big] + \omega^2a^3b^3\bigr\},
\end{split}
\end{equation}
for type II (with $\omega ab>\frac{b-a}{a}$), and
\begin{equation}
\begin{split}
  {\cal N}_1(x) &= - \frac{8}{ab(b-a)^3} \bigl\{-\omega ab\big[\omega ab(a+b)-(b-a)\big]\big[\omega ab(a+b)
       -2(b-a)\big]x \\
  &\quad + \big[\omega ab(a+b)-2(b-a)\big](b-a)^2 + \omega^2a^2b^2(b-a)^3\bigr\}, \\
  {\cal N}_2(x) &= \frac{128\omega}{(b-a)^5}\big[\omega ab(a+b) - 2(b-a)\big]\bigl\{\omega^2a^3b^3
       - (b-a)\big[\omega ab(a+b) - (b-a)\big]\bigr\} \\
  &\quad \times \bigl\{-\big[\omega ab(a+b) - (b-a)\big]x + \omega a^2b^2\bigr\}, \\
  {\cal D}(x) &= \frac{4}{(b-a)^5}\bigl\{\big[\omega ab(a+b) - (b-a)\big]\big[\big(\omega ab(a+b) - 2(b-a)\big)x
       - 2\omega a^2b^2  \\
  &\quad + 2a(b-a)\big]\big[\big(\omega ab(a+b)-2(b-a)\big) x - 2\omega a^2b^2 + 2b(b-a)\big] \\
  &\quad + \omega^2a^3b^3(b-a)^3\bigr\},
\end{split}
\end{equation}
for type III (with $\omega ab>\frac{b-a}{a}$). The corresponding spectrum is given by (\ref{eq:E}), where $n$ runs over $n=0$, 1, 2,\ldots\ in the type I and II cases, and over $n=-2$, 1, 2,\ldots\ in the type III one.\par
%
%
As examples, for $\omega=a=1$, $b=3$, we obtain
\begin{equation}
\begin{split}
  V_{\rm eff,ext}(x) &= \frac{3x^2}{4(x-1)(3-x)} + \frac{16(3x-23)}{3(2x^2+16x-3)} + \frac{280(8x-3)}
       {(2x^2+16x-3)^2}, \\
  V_{\rm eff,ext}(x) &= \frac{3x^2}{4(x-1)(3-x)} + \frac{8(60x-59)}{15(20x^2-32x+15)} - \frac{88(16x-15)}
       {5(20x^2-32x+15)^2}, \\
  V_{\rm eff,ext}(x) &= \frac{3x^2}{4(x-1)(3-x)} + \frac{2(30x-13)}{3(20x^2-50x+33)} - \frac{14(10x-9)}
       {(20x^2-50x+33)^2},
\end{split}
\end{equation}
for type I, II, and III, respectively.\par
%
%
\section{Conclusion}

In the present paper, we have first shown that the PDM model of Ref.~\cite{jafarov} can be alternatively solved by applying a PCT to the constant-mass Schr\"odinger equation for the Scarf I potential.\par
%
%
In a second step, on starting from the known rational extension of the latter connected with $X_1$-Jacobi EOPs, we have built a rationally-extended PDM model with the same spectrum as the starting model and wavefunctions expressed in terms of $X_1$-Jacobi EOPs instead of Jacobi polynomials.\par
%
%
{}Finally, we have sketched how some more involved PDM models may be obtained by starting from the rationally-extended Scarf I potentials related to $X_2$-Jacobi EOPs.\par
%
%
\section*{Data availability statement}

No new data were created or analyzed in this study.\par
%
%
\section*{Acknowledgment}

The author was supported by the Fonds de la Recherche Scientifique-FNRS under Grant No.~4.45.10.08.\par
%
%

\end{document}